\documentclass[pdflatex,sn-basic]{sn-jnl}

\jyear{2022}%

\theoremstyle{thmstyleone}%
\usepackage{xcolor}
\usepackage{graphicx}
\usepackage{amsmath}
\usepackage{amssymb}
\usepackage{booktabs}
\usepackage{subcaption}
\usepackage{multirow}
\usepackage[export]{adjustbox}

\theoremstyle{thmstyletwo}%

\theoremstyle{thmstylethree}%

\begin{document}


\title[Efficient multi-scale visual object representation]{Efficient multi-scale representation of visual objects using a biologically plausible spike-latency code and winner-take-all inhibition}


\author[1]{\fnm{{\normalsize Melani}} \sur{\normalsize Sanchez-Garcia}}\email{{\small mesangar@ucsb.edu}}
\equalcont{MSG and TC are co-first authors. BRC and MB are co-last authors.}

\author[2,3]{\fnm{\normalsize Tushar} \sur{Chauhan}}\email{{\small tchauhan@mit.edu}}
\equalcont{MSG and TC are co-first authors. BRC and MB are co-last authors.}

\author[3,4]{\fnm{\normalsize Benoit R. } \sur{Cottereau}}\email{{\small benoit.cottereau@cnrs.fr}}
\equalcont{MSG and TC are co-first authors. BRC and MB are co-last authors.}

\author[1,5]{\fnm{\normalsize Michael} \sur{Beyeler}}\email{{\small mbeyeler@ucsb.edu}}
\equalcont{MSG and TC are co-first authors. BRC and MB are co-last authors.}

\affil[1]{\orgdiv{{\small Department of Computer Science}}, \orgname{{\small University of California}}, \orgaddress{\city{{\small Santa Barbara}}, \state{{\small CA}}, \country{{\small USA}}}}

\affil[2]{\orgdiv{{\small The Picower Institute for Learning and Memory}}, \orgname{{\small Department of Brain and Cognitive Sciences, Massachusetts Institute of Technology}}, \orgaddress{\city{{\small Boston}}, \state{{\small MA}}, \country{{\small USA}}}}

\affil[3]{\orgdiv{{\small CerCo CNRS UMR5549}}, \orgname{{\small Universit\'{e} de Toulouse III-Paul Sabatier, Toulouse}}, \orgaddress{\country{{\small France}}}}

\affil[4]{\orgdiv{{\small IPAL}}, \orgname{{\small CNRS IRL 2955}}, \orgaddress{\country{{\small Singapore}}}}

\affil[5]{\orgdiv{{\small Department of Psychological \& Brain Sciences}}, \orgname{{\small University of California}}, \orgaddress{\city{{\small Santa Barbara}}, \state{{\small CA}}, \country{{\small USA}}}}


\abstract{Deep neural networks have surpassed human performance in key visual challenges such as object recognition, but require a large amount of energy, computation, and memory. 
In contrast, spiking neural networks (SNNs) have the potential to improve both the efficiency and biological plausibility of object recognition systems.
Here we present a SNN model that uses spike-latency coding and winner-take-all inhibition (WTA-I) to efficiently represent visual stimuli using multi-scale parallel processing.
Mimicking neuronal response properties in early visual cortex, images were preprocessed with three different spatial frequency (SF) channels,
before they were fed to a layer of spiking neurons whose synaptic weights were updated using spike-timing-dependent-plasticity (STDP).
We investigate how the quality of the represented objects changes under different SF bands and WTA-I schemes.
We demonstrate that a network of 200 spiking neurons tuned to three SFs can efficiently represent objects with as little as 15 spikes per neuron. 
Studying how core object recognition may be implemented using biologically plausible learning rules in SNNs may not only further our understanding of the brain, but also lead to novel and efficient artificial vision systems.}

\keywords{spiking neural networks, spike-timing-dependent-plasticity, multi-scale processing, spike-latency code, winner-take-all inhibition}



\maketitle 
\section{Introduction} 
\label{sec1}
Deep convolutional neural network (DCNNs) have been extremely successful in a wide range of computer vision applications, rivaling or exceeding human benchmark performance in key visual challenges such as object and face recognition \citep{he2015delving,sun2015deepid3} or scene categorization \citep{stivaktakis2019deep}. However, state-of-the-art DCNNs require too much energy, computation, and memory to be deployed on most computing devices and embedded systems \citep{goel2020survey}. In contrast, the brain is masterful at representing real-world objects with a cascade of reflexive, largely feedforward computations \citep{dicarlo2012does} that rapidly unfold over time \citep{ales_time_2013,cichy2016similarity} and rely on an extremely sparse, efficient neural code (for a recent review see \citet{beyeler2019neural}). For example, in macaques, faces are processed in localized patches along the Superior Temporal Sulcus (STS), where cells detect distinct constellations of face parts (e.g., eyes, noses, mouths), and whole faces can be recognized from a linear combination of neural responses within these face patches \citep{chang2017code,majaj2015simple}. 

In recent years, spiking neural networks (SNNs) have emerged as a promising approach to improving the efficiency and biological plausibility of neural networks such as DCNNs, due to their potential for low power consumption, fast inference, event-driven processing, and asynchronous operation \citep{gerstner2002spiking}.
To facilitate learning in such networks, new learning algorithms based on varying degrees of biological plausibility have also been developed recently. 
For instance, spike-timing-dependent plasticity (STDP) is an unsupervised learning rule that is observed in biological systems \citep{bi1998synaptic,caporale2008spike} and that can be used to extract the most notable spike patterns \citep{feldman2012spike,brzosko2019neuromodulation} by adjusting the efficacy of synaptic connections based on the relative timing of presynaptic and postsynaptic spikes.
Studying how object recognition may be implemented using biologically plausible learning rules in SNNs may not only further our understanding of the brain, but also lead to the development of energy efficient systems, implementable on neuromorphic hardware.

Here we present a SNN model that uses spike-latency coding \citep{chauhan2018emergence,chauhan2021sub} and winner-take-all inhibition (WTA-I) \citep{maass2000computational} to efficiently represent visual stimuli using multi-scale parallel processing. 
Part of this work \citep{sanchez2022efficient} was previously presented at the CVPR'22 NeuroVision workshop\footnote[1]{\url{https://sites.google.com/uci.edu/neurovision2022}}. 
Given an input image, stimuli were preprocessed with parallel spatial frequency (SF) channels mimicking the sensitivity of neurons in early visual cortex \citep{de1982spatial}.
The resulting combination of the SF channels was then fed to a layer of spiking neurons whose synaptic weights were updated using STDP \citep{gutig2003learning}.
We show that STDP can learn efficient object representations from the MNIST \citep{lecun1998mnist}, FASHION-MNIST \citep{xiao2017fashion}, CIFAR10 \citep{krizhevsky2009learning}, and ORL \citep{samaria1994parameterisation} datasets. In addition, we investigate how the quality of the represented objects changes under different SF bands and WTA-I schemes.
Remarkably, our network is able to represent objects with as little as 200 neurons and 15 spikes per neuron.


The rest of the paper is organized as follows: Section~\ref{sec2} briefly introduces some of the most recent related works. Section~\ref{sec3} explains the main framework and the model equations. Next, we report the results of a computational study in which we explored the quality of the represented objects and the sparsity trade-off for the different networks schemes (see Section~\ref{sec4}). Finally, a brief Discussion summarizes the main results and gives some perspectives in Section~\ref{sec5}.

\section{Related Work} \label{sec2}
Significant efforts have been expended in recent years to demonstrate the efficacy of SNNs with STDP in object recognition applications. Previous studies have used STDP to extract visual features of low or intermediate complexity from images and without supervision.
\citet{yu2013rapid} proposed a novel SNN with a supervised learning rule and temporal coding scheme to generate temporal spike patterns, which could be used to classify a subset of handwritten digits found in the MNIST database.
\citet{liu2016visual} combined Gabor filter banks with rank-order coding and STDP to push the MNIST classification rate to 82\%.
\citet{beyeler_categorization_2013} achieved 92\% on MNIST using a Calcium-based STDP learning rule.
\citet{masquelier2007unsupervised} used the STDP rule in an asynchronous feedforward SNN that mimics the ventral visual pathway and showed the emergence of selectivity to intermediate-complexity visual features when the network was presented with natural images. 

More recent articles designed a deep SNN, comprising several convolutional and pooling layers trainable with either standard STDP \citep{kheradpisheh2018stdp} or reward-based STDP \citep{mozafari2019bio}. Studying how object recognition may be implemented using biologically plausible learning rules in SNNs may not only further our understanding of the brain, but also lead to new efficient artificial vision systems. 

Theories on visual perception claim the existence of multiple channels, or multiple receptive field (RF) sizes, in the early visual processing and the importance of the spatial frequency (SF) contents of images during object recognition \citep{kauffmann2014neural, ginsburg1986spatial, field1987relations, tolhurst1992amplitude,hughes1996global}. Because RFs of neuronal populations in the visual pathway vary in size, the responses of different subsets of neurons would constitute a neural representation at some particular scale,
allowing us to represent visual scenes as a combination of SF channels
\citep{campbell1973transmission}.

\begin{figure}[t]
    \centering
    \includegraphics[width=\linewidth]{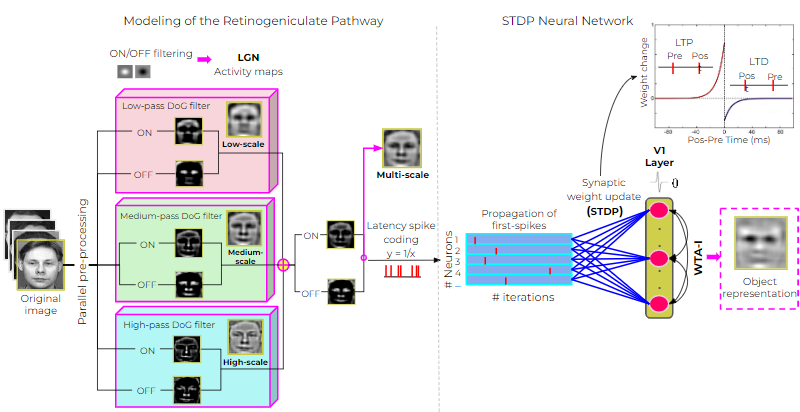}
    \caption{Multi-scale network, illustrated using images from the ORL dataset \citep{samaria1994parameterisation}.
    Images were convolved with ON and OFF center/surround kernels to simulate LGN responses. To simulate the multiple channels in the visual system, we used a pre-processing scheme where LGN maps are generated based on a particular SF range: Low-scale, Medium-scale and High-scale (further illustrated in Fig.~\ref{fig:preprocessing}). The three LGN responses were added, converted to spike latencies, and fed to a spiking neural network (SNN) with plastic synapses implementing spike-timing-dependent-plasticity (STDP) and winner-take-all inhibition (WTA-I). The propagated LGN spikes contributed to an increase in the membrane potential of V1 neurons until one of the V1 membrane potentials reached threshold, resulting in a postsynaptic spike and inhibition of all other V1 neurons until the next iteration. The synaptic weights were updated using an unsupervised STDP rule. Objects were reconstructed by taking a linear combination of spiking activity across the V1 population. }
    \label{fig:overview}
\end{figure}


Selectivity for SF is one of the fundamental and most thoroughly studied properties of visual neurons \citep{henriksson2008spatial,shapley1985spatial,de_valois_spatial_1982}. The primary visual system processes low-level and high-level stimulus properties using inputs from the retina via the lateral geniculate nucleus (LGN). 
In the earliest stages of the visual pathway, the processing of different stimulus attributes occurs in a parallel fashion. This means that images are filtered by parallel, SF-selective channels \citep{enroth1966contrast}, which may converge in V1 \citep{nassi2009parallel}. The visual information from the LGN passes through V1 and multiple strategies might be used to transfer parallel input into multiple output streams. 


\section{Methods}\label{sec3}


\begin{figure}[t]
    \centering
    \includegraphics[width=0.9\linewidth]{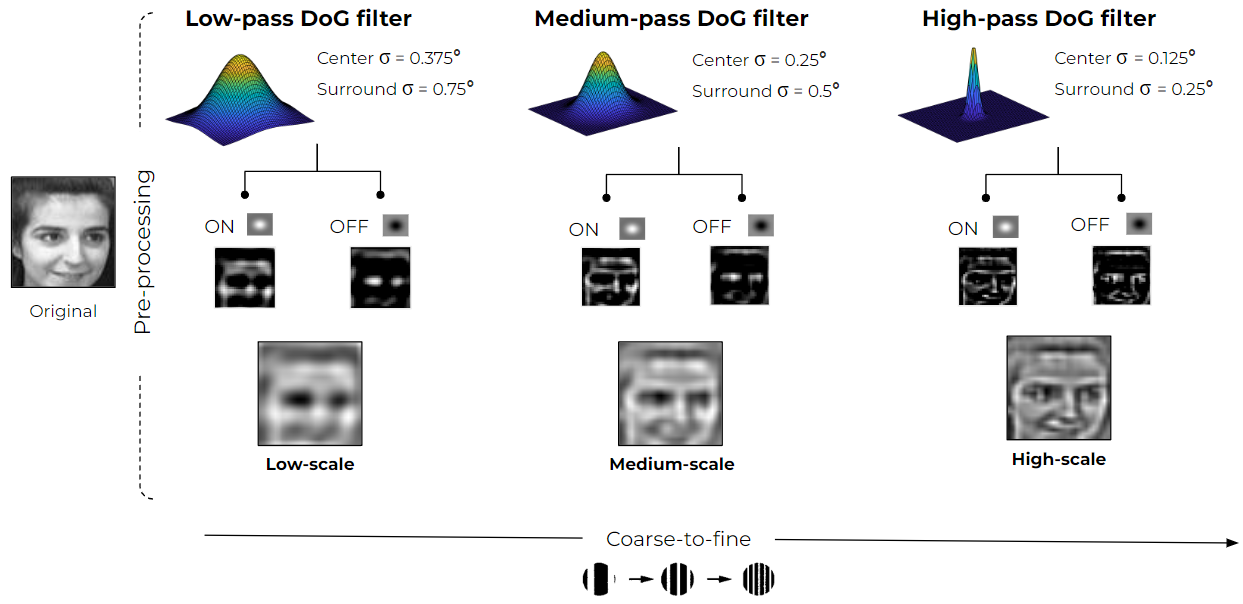}
    \caption{LGN preprocessing. To simulate the computations performed by the retinal ganglion cells and the LGN, the images were convolved with ON and OFF center-surround kernels \citep{chauhan2018emergence}. Specifically, we chose three sizes based on an earlier study \citep{chauhan2018emergence}:
    0.375$^{\circ}$/0.75$^{\circ}$ for low SF, 0.25$^{\circ}$/0.5$^{\circ}$ for medium SF and 0.125$^{\circ}$/0.25$^{\circ}$ for high SF \citep{solomon2002extraclassical}. The resulting images processed with these filters correspond to Low-scale, Medium-scale and High-scale LGN maps, respectively.}
    \label{fig:preprocessing}
\end{figure}


\begin{figure}[t]
    \centering
    \includegraphics[width=0.8\linewidth]{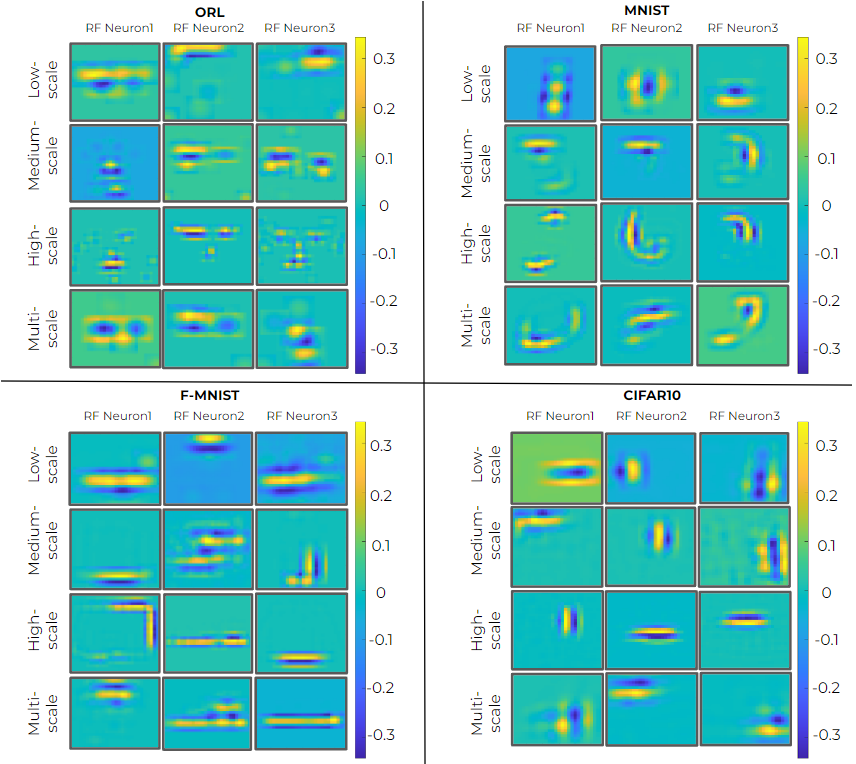}
    \caption {Example RFs of three representative neurons (columns in each panel) of the simulated population for low-scale, medium-scale, high-scale and multi-scale networks (rows). With STDP, neurons progressively learned features corresponding to prototypical patterns that were both salient and frequent.}
    \label{fig:rfs}
\end{figure}


\subsection{Network architecture}

\noindent The network architecture of our model is shown in Fig.~\ref{fig:overview}.
Inspired by \citet{chauhan2018emergence}, our network consisted of an input layer corresponding to a simplified model of the LGN, followed by a layer of spiking neurons whose synaptic weights were updated using STDP. 
The LGN layer consisted of simulated firing-rate neurons with center-surround RFs, implemented using DoG filters which simulate the computations performed by the retinal ganglion cells and the LGN (\citet{enroth1966contrast,derrington1982influence}; further illustrated in Fig.~\ref{fig:preprocessing}). 
Based on \citet{chauhan2018emergence}, the RF sizes were chosen to reflect the size of representative LGN center-surround magnocellular RFs. 
It is well known that the SFs of LGN cells differ by about a factor of 3; meaning that some cells are most sensitive to patterns that contain relatively high SFs, whereas other cells are most sensitive to patterns of low SFs \citep{derrington1979mechanism}. 
Specifically, we chose three sizes of center-surround RFs within the range of SFs for a magnocellular cell: 0.375$^{\circ}$/0.75$^{\circ}$ for low SF, 0.25$^{\circ}$/0.5$^{\circ}$ for medium SF and 0.125$^{\circ}$/0.25$^{\circ}$ for high SF \citep{solomon2002extraclassical}. These values correspond to the widths of the gaussian used for the DoG filter. The resulting LGN images processed with these filters corresponded to low-scale, medium-scale and high-scale images (see left-hand side of Fig.~\ref{fig:preprocessing}).
The three LGN responses were added and converted to spike latencies \citep{chauhan2018emergence}. 
The LGN layer was fully connected to a layer of integrate-and-fire neurons, each unit characterized by a threshold and a membrane potential \citep{chauhan2018emergence}. 
The LGN spikes contributed to an increase in the membrane potential of V1 neurons, until one of the V1 membrane potentials reached threshold, resulting in a postsynaptic spike.
The proposed method is compared with an alternative network architecture, which can be found in Appendix~\ref{secA1}.

\subsection{Neuron model}
The membrane potential $E_{n}(t)$ of the $n$-th V1 neuron at time $t$ within the iteration was represented as:
\begin{equation}
\resizebox{0.9\hsize}{!}{
${E_{n}(t) =\begin{cases}  \sum\limits_{m\in LGN} w_{mn}\cdot H(t-t_{m}), \,\,\,\, & t < \min\limits_{t} \Big\{ t \mid \max\limits_{n \in V1} E_{n}(t) \ge \theta \Big\}  \\
          &         0, \,\,\,\, \text{otherwise}. \end{cases}}$}
    \label{eq:1}
\end{equation}
where $t_{m}$ was the spike time of the $m$-th LGN neuron, ${H}$ was the Heaviside or unit step function, and $\theta$ was the threshold of the V1 neurons (assumed to be a constant shared by the entire population). The expression $\min \{ t \mid \max E_{n}(t) \ge \theta \}$ denoted the timing of the first spike in the V1 layer.
Membrane potentials were calculated up to this point in time, after which a WTA-I scheme \citep{maass2000computational} was triggered and all membrane potentials were reset to zero.
In this scheme, the most frequently firing neuron exerted the strongest inhibition on its competitors and thereby stopped them from firing until the end of the iteration.

\subsection{Spike-latency code}

\noindent Following \citet{chauhan2018emergence}, we converted the LGN activity maps to first-spike relative latencies using a simple inverse operation: $y = 1/x$, where $x$ was the LGN input and $y$ was the assigned spike-time latency. Any monotonically decreasing function would lead to equivalent results (i.e., where the most active
units fire first, while units with lower activity fire later or not at all) (see \citep{masquelier2007unsupervised}). In this way, we ensured that the most active units fired first, while units with lower activity fired later or not at all.

\subsection{Spike-timing-dependent-plasticity}

\noindent The weights of plastic synapses connecting LGN and V1 were updated using multiplicative STDP, which is an unsupervised learning rule that modifies synaptic strength, $w$, as a function of the relative timing of pre- and postsynaptic spikes, $\Delta t$ \citep{gutig2003learning}. LTP ($\Delta{t} > 0$) and LTD ($\Delta{t} \le 0$) were driven by their respective learning rates $\alpha^{+}$ and $\alpha^{-}$, leading to a weight change ($\Delta w$):
\begin{equation}
\Delta{w}=\begin{cases} -\alpha^{-} \cdot w^{\mu^{-}} \cdot K(\Delta{t},\tau_{-}), \Delta{t} \le 0 \\
                     \alpha^{+} \cdot (1-w)^{\mu^{+}} \cdot K(\Delta{t},\tau_{+}), \Delta{t} > 0,
       \end{cases} \label{eq:2}
\end{equation}
where $\alpha^{+} = 5 \times 10^{-3}$ and $\alpha^{-} = 3.75 \times 10^{-3}$, $K(\Delta{t}, \tau) = e^{-\vert\Delta{t}\vert/\tau}$ was a temporal windowing filter, and $\mu^{+} = 0.65$ and $\mu^{-} = 0.05$ were constants $\in [0, 1]$ that defined the nonlinearity of the LTP and LTD process, respectively. 
STDP has the effect of concentrating high synaptic weights on afferents that systematically fire early, thereby decreasing postsynaptic spike latencies for these connections.

In this implementation, computation speed greatly increased by making the windowing filter $K$ infinitely wide, which is equivalent to assuming $\tau_{\pm} \to \infty$ or $K = 1$ \citep{gutig_learning_2003}.
A ratio $\alpha^{+} /\alpha^{-} = 4/3$ was chosen based on previous experiments that demonstrated network stability \citep{masquelier2007unsupervised}. Also, \citet{chauhan2018emergence} showed that the results were robust to variations of this ratio. The threshold of the V1 neurons was fixed through trial and error at $\theta = 20$. This value was unmodified for all experiments.

Initial weight values were sampled from a random uniform distribution between 0 and 1.
After each iteration, the synaptic weights for the first V1 neuron to fire were updated using STDP (Equation~\ref{eq:2}), and the membrane potentials of all the other neurons in the V1 population were reset to zero. 
The STDP rule was active only during the training phase.

\begin{figure}[!t]
     \centering
     \begin{subfigure}[b]{0.45\textwidth}
         \centering
         \includegraphics[width=\textwidth]{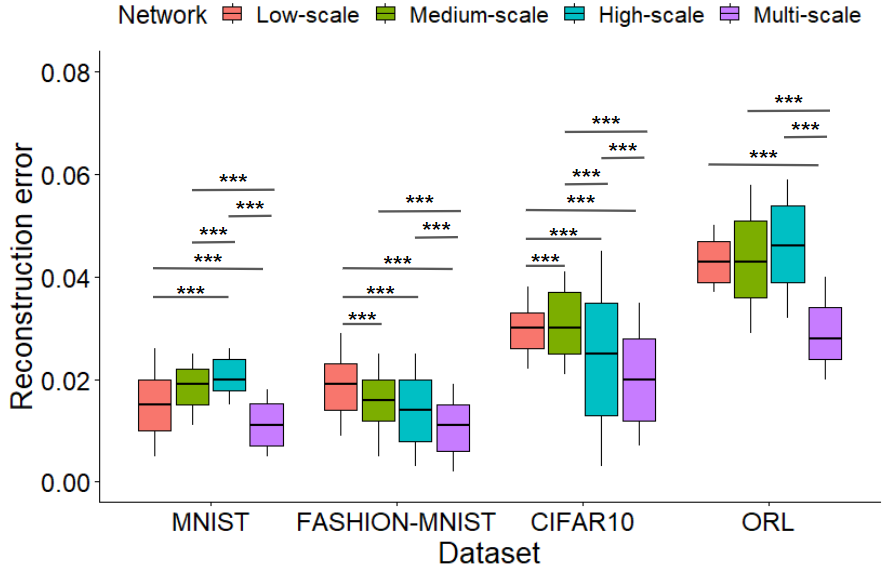}
         \caption[a]{}
     \end{subfigure}
     \begin{subfigure}[b]{0.45\textwidth}
         \centering
         \includegraphics[width=\textwidth]{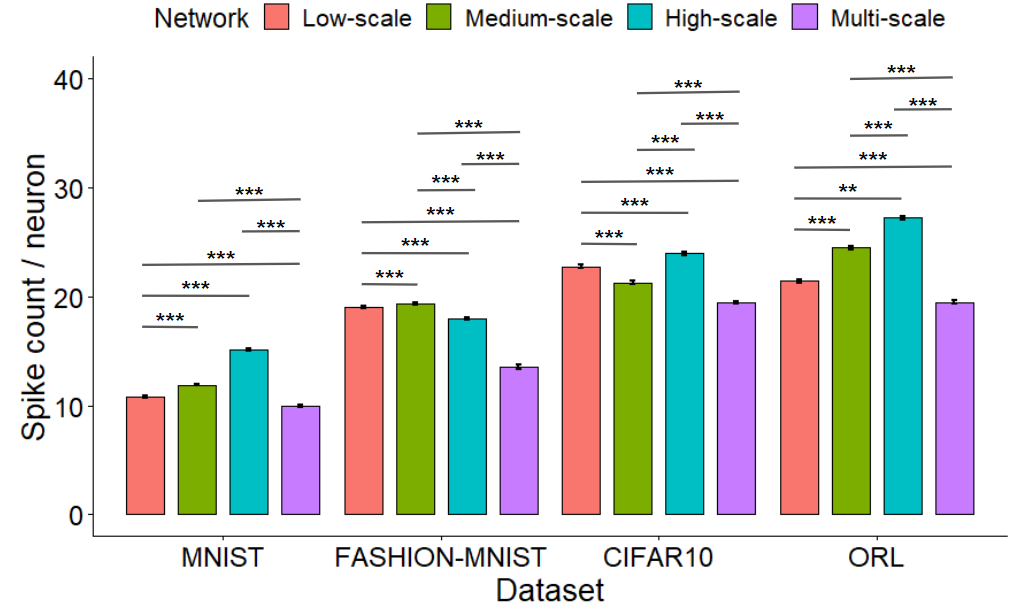}
         \caption{}
     \end{subfigure}
     \begin{subfigure}[b]{0.45\textwidth}
         \centering
         \includegraphics[width=\textwidth]{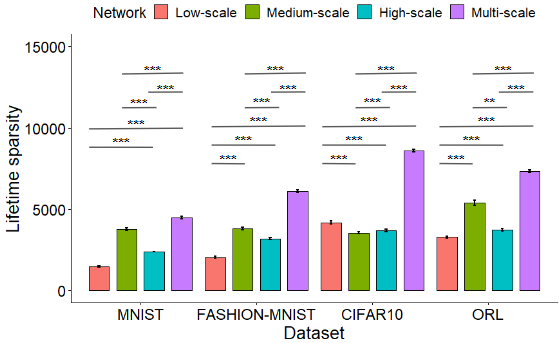}
         \caption{}
     \end{subfigure}
     \begin{subfigure}[b]{0.45\textwidth}
         \centering
         \includegraphics[width=\textwidth]{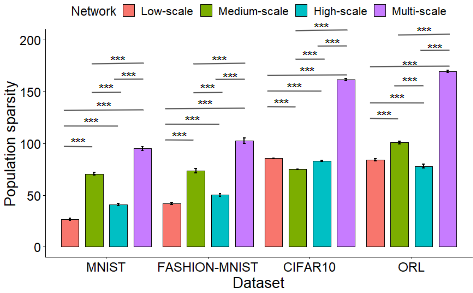}
         \caption{}
     \end{subfigure}
        \caption{Multi-scale network. (a) Reconstruction error of test set. (b) Spike count per neuron: number of spikes fired by an active neuron. (c) Lifetime sparsity: active stimuli during the lifetime of a neuron. (d) Population sparsity: neurons active at any point in time. Mean responses and standard deviation grouped by type of network (Low-scale, Medium-scale, High-scale and Multi-scale). Error bars have been averaged across neurons for lifetime sparsity and averaged across images for population sparsity. $*** = p<.001$; $** = p<.01$; $* = p<.05$; $ns = p>.05$. All t-tests paired samples, two-tailed.}

        \label{fig:graph1}
\end{figure}


\begin{table}[]
\caption{Global results for type of networks. Comparison of mean responses and standard deviation grouped by type of network and dataset.}\label{tab1}%
 \begin{adjustbox}{width=\textwidth}
\begin{tabular}{|c|c|c|c|c|c|}
\toprule
{Dataset} & {Network} & \footnotesize RE   & \footnotesize LS   & \footnotesize PS   & \footnotesize SC  \\ 
\midrule
\multirow{4}{*}{MNIST} & \footnotesize Low-scale   &  1.81e-2$\pm$4.53e-3   &   1483.7$\pm$695.0   & 26.6$\pm$11.9     &  10.8$\pm$0.77  \\ 
& \footnotesize Medium-scale   &   2.07e-2$\pm$1.29e-2   &  3788.0$\pm$103.2    &  70.6$\pm$5.14    &  11.9$\pm$0.78  \\ 
& \footnotesize High-scale     &  1.51e-2$\pm$4.84e-3    &  2386.7$\pm$295.9    &  40.9$\pm$3.31    &  15.1$\pm$0.81   \\
& \footnotesize Multi-scale     &  1.16e-2$\pm$3.77e-3    &  4500.1$\pm$782.7    &  95$\pm$19.9    &   10$\pm$0.79  \\

\midrule
\multirow{4}{*}{FASHION-MNIST} & \footnotesize Low-scale   &  1.49e-2$\pm$5.99e-3      &   2037.5$\pm$735.4     &  42.1$\pm$19.8      &   19.0$\pm$1.07 \\ 
& \footnotesize Medium-scale   &   1.37e-2$\pm$6.87e-3     &    3822.9$\pm$493.8    &  73.5$\pm$11.12      &  19.3$\pm$1.39   \\ 
& \footnotesize High-scale   &   1.90e-2$\pm$6.17e-3     &   3201.8$\pm$591.0     &  50.1$\pm$10.39      &  18.0$\pm$1.38   \\
& \footnotesize Multi-scale  &   9.34e-3$\pm$5.15e-3     &  6105.0$\pm$907.9      &   102.5$\pm$25.2     &   13.6$\pm$1.76    \\

\midrule
\multirow{4}{*}{CIFAR10} & \footnotesize Low-scale & 3.10e-2$\pm$6.66e-3     &   4179.9$\pm$795.7    &  85.5$\pm$3.29    &  22.8$\pm$1.72  \\ 
& \footnotesize Medium-scale    &   2.13e-2$\pm$3.43e-3    &  3542.4$\pm$693.9     &  75.0$\pm$11.18    &  21.3$\pm$1.40  \\ 
& \footnotesize High-scale     &   3.07e-2$\pm$6.61e-3    &   3692.9$\pm$1006.7    &  83.2$\pm$8.32    &  24.0$\pm$1.75 \\
& \footnotesize Multi-scale     &  2.15e-2$\pm$8.22e-3    &  8599.5$\pm$830.7     &  161.5$\pm$10.8    &  19.5$\pm$1.06  \\

\midrule
\multirow{4}{*}{ORL} & \footnotesize Low-scale   &  4.54e-2$\pm$8.40e-3   &   3282.5$\pm$1525.4   &  84.0$\pm$11.97   &  21.4$\pm$1.08   \\ 
& \footnotesize Medium-scale   &  4.54e-2$\pm$8.43e-3    &   5404.5$\pm$704.3   & 100.8$\pm$18.32    & 24.5$\pm$1.48    \\ 
& \footnotesize High-scale     &   4.30e-2$\pm$3.99e-3   &  3732.7$\pm$559.5    &  78.0$\pm$9.24   &  27.2$\pm$1.72  \\
& \footnotesize Multi-scale    &  2.91e-2$\pm$6.07e-3    &  7320$\pm$847.0    &  169.4$\pm$11.7   &  19.5$\pm$1.78  \\

\bottomrule
\end{tabular}
\end{adjustbox}
\end{table}


\begin{figure}[!t]
\centering
\begin{tabular}{c c c c c c }
& {\footnotesize Low-scale} & {\footnotesize Medium-scale} & {\footnotesize High-scale} & {\footnotesize Multi-scale} \\

\rotatebox[origin=c]{90}{{\footnotesize ORL}} \rotatebox[origin=c]{90} {\footnotesize
 Target }& \raisebox{-0.5\height}{\includegraphics[width=0.12\textwidth]{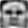}} &
\raisebox{-0.5\height}{\includegraphics[width=0.12\textwidth]{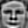}} &
\raisebox{-0.5\height}{\includegraphics[width=0.12\textwidth]{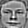}} &
\raisebox{-0.5\height}{\includegraphics[width=0.12\textwidth]{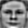}}\\
\rotatebox[origin=c]{90}{{\footnotesize ORL}} \rotatebox[origin=c]{90} {\footnotesize
 Prediction }& \raisebox{-0.5\height}{\includegraphics[width=0.12\textwidth]{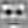}} &
\raisebox{-0.5\height}{\includegraphics[width=0.12\textwidth]{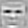}} &
\raisebox{-0.5\height}{\includegraphics[width=0.12\textwidth]{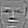}} &
\raisebox{-0.5\height}{\includegraphics[width=0.12\textwidth, cfbox=black 2pt 1pt]{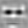}}\\
&  {\footnotesize 0.04216} & {\footnotesize 0.04408} & {\footnotesize 0.04567} & {\footnotesize 0.03772} \\

\rotatebox[origin=c]{90}{ {\footnotesize
 MNIST }} \rotatebox[origin=c]{90} {\footnotesize
 Target } & \raisebox{-0.5\height}{\includegraphics[width=0.12\textwidth]{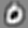}} &
\raisebox{-0.5\height}{\includegraphics[width=0.12\textwidth]{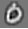}} &
\raisebox{-0.5\height}{\includegraphics[width=0.12\textwidth]{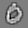}} & \raisebox{-0.5\height}{\includegraphics[width=0.12\textwidth]{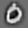}} \\
\rotatebox[origin=c]{90}{{\footnotesize MNIST}} \rotatebox[origin=c]{90} {\footnotesize
 Prediction } & \raisebox{-0.5\height}{\includegraphics[width=0.12\textwidth]{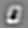}} &
\raisebox{-0.5\height}{\includegraphics[width=0.12\textwidth]{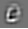}} &
\raisebox{-0.5\height}{\includegraphics[width=0.12\textwidth]{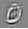}} &
\raisebox{-0.5\height}{\includegraphics[width=0.12\textwidth, cfbox=black 2pt 1pt]{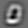}}\\
&  {\footnotesize 0.01656} & {\footnotesize 0.01716} & {\footnotesize 0.01854} & {\footnotesize 0.01057} \\

\rotatebox[origin=c]{90}{{\footnotesize F-MNIST}} \rotatebox[origin=c]{90} {\footnotesize
 Target }  & \raisebox{-0.5\height}{\includegraphics[width=0.12\textwidth]{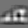}} &
\raisebox{-0.5\height}{\includegraphics[width=0.12\textwidth]{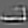}} &
\raisebox{-0.5\height}{\includegraphics[width=0.12\textwidth]{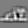}} &
\raisebox{-0.5\height}{\includegraphics[width=0.12\textwidth]{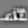}}\\
\rotatebox[origin=c]{90}{{\footnotesize F-MNIST}} \rotatebox[origin=c]{90} {\footnotesize
 Prediction }  & \raisebox{-0.5\height}{\includegraphics[width=0.12\textwidth]{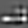}} &
\raisebox{-0.5\height}{\includegraphics[width=0.12\textwidth]{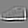}} &
\raisebox{-0.5\height}{\includegraphics[width=0.12\textwidth]{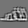}} &
\raisebox{-0.5\height}{\includegraphics[width=0.12\textwidth, cfbox=black 2pt 1pt]{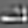}}\\
&  {\footnotesize 0.01737} & {\footnotesize 0.01770} & {\footnotesize 0.01876} & {\footnotesize 0.01019} \\

\rotatebox[origin=c]{90}{{\footnotesize CIFAR10}} \rotatebox[origin=c]{90} {\footnotesize
 Target }& \raisebox{-0.5\height}{\includegraphics[width=0.12\textwidth]{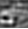}} &
\raisebox{-0.5\height}{\includegraphics[width=0.12\textwidth]{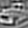}} &
\raisebox{-0.5\height}{\includegraphics[width=0.12\textwidth]{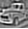}} &
\raisebox{-0.5\height}{\includegraphics[width=0.12\textwidth]{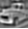}}\\
\rotatebox[origin=c]{90}{{\footnotesize CIFAR10}} \rotatebox[origin=c]{90} {\footnotesize
 Prediction }& \raisebox{-0.5\height}{\includegraphics[width=0.12\textwidth]{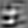}} &
\raisebox{-0.5\height}{\includegraphics[width=0.12\textwidth]{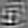}} &
\raisebox{-0.5\height}{\includegraphics[width=0.12\textwidth]{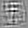}} &
\raisebox{-0.5\height}{\includegraphics[width=0.12\textwidth, cfbox=black 2pt 1pt]{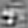}}\\
&  {\footnotesize 0.02946} & {\footnotesize 0.03689} & {\footnotesize 0.03099} & {\footnotesize 0.02068} \\

\end{tabular}
\caption{Representative object representation (OR) examples using low-scale, medium-scale, high-scale and multi-scale networks (columns). 
The number below each image indicates the reconstruction error for that particular image. The black frame highlights the image with the smallest error.}
\label{fig:reconstructedImages}
\end{figure}

\begin{figure}[!t]
     \centering
     \begin{subfigure}[b]{0.45\textwidth}
         \centering
         \includegraphics[width=\textwidth]{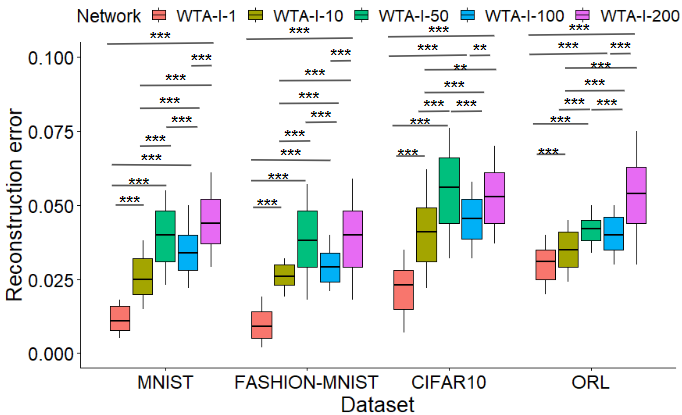}
         \caption[a]{}
     \end{subfigure}
     \begin{subfigure}[b]{0.45\textwidth}
         \centering
         \includegraphics[width=\textwidth]{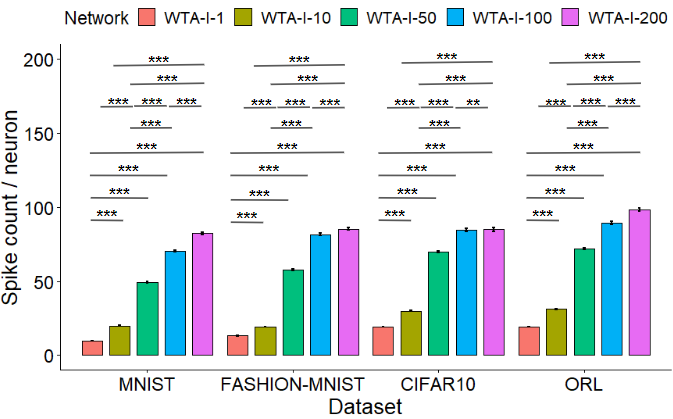}
         \caption{}
     \end{subfigure}
     \begin{subfigure}[b]{0.45\textwidth}
         \centering
         \includegraphics[width=\textwidth]{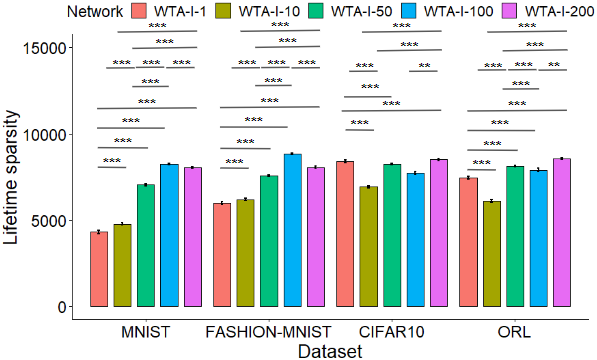}
         \caption{}
     \end{subfigure}
     \begin{subfigure}[b]{0.45\textwidth}
         \centering
         \includegraphics[width=\textwidth]{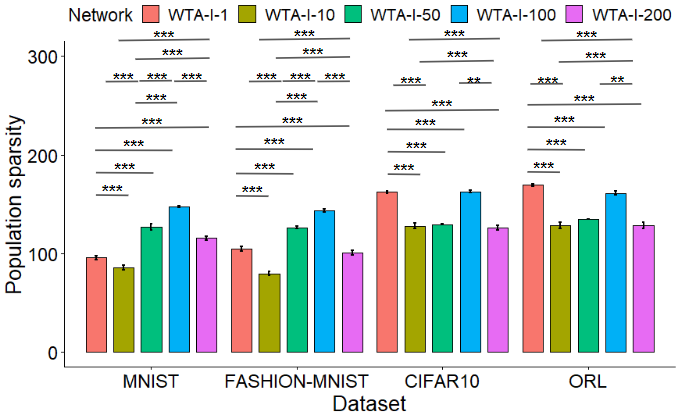}
         \caption{}
     \end{subfigure}
        \caption{WTA-I schemes. (a) Reconstruction error in the test phase as a function of the number of spikes included in the STDP algorithm (WTA-I) for 200 V1 neurons. (b) Lifetime sparsity: active stimuli during the lifetime of a neuron. (c) Population sparsity: neurons active at any point in time. (d) Spike count per neuron: number of spikes fired by an active neuron. Mean responses and standard deviation grouped by the WTA-I schemes. Error bars have been averaged across neurons for lifetime sparsity and averaged across images for population sparsity. }

        \label{fig:WTAISpikes}
\end{figure}


\begin{table}[]
\caption{Global results for WTA-I schemes. Comparison of mean responses and standard deviation grouped by type of WTA-I schemes and dataset.}\label{}%
 \begin{adjustbox}{width=\textwidth}
\begin{tabular}{|c|c|c|c|c|c|}
\toprule
{Dataset} & {WTA-I} & \footnotesize RE   & \footnotesize LS   & \footnotesize PS   & \footnotesize SC  \\ 
\midrule

\multirow{4}{*}{MNIST} & \footnotesize WTA-I-1    &  1.16e-2$\pm$3.77e-3    &  4500.1$\pm$782.7    &  95.0$\pm$19.9    &   10.0$\pm$0.8   \\
& \footnotesize WTA-I-10  &  1.15e-2$\pm$4.33e-3    &  4319.9$\pm$773.7    &  95.7$\pm$22.4    &  9.8$\pm$0.8 \\
& \footnotesize WTA-I-50  &  2.67e-2$\pm$1.18e-2    &  4695.9$\pm$536.7    &  87.5$\pm$25.4    &   20.1$\pm$1.3  \\
& \footnotesize WTA-I-100  & 3.41e-2$\pm$7.48e-3    &  7919.3$\pm$558.42   & 109.2$\pm$28.3   &   49.9$\pm$5.5 \\
& \footnotesize WTA-I-200  &  3.93e-2$\pm$9.61e-3    &  7094.6$\pm$458.8    &  128.9$\pm$28.1    &   70.1$\pm$5.4 \\

\midrule
\multirow{4}{*}{FASHION-MNIST}
& \footnotesize WTA-I-1  &   9.34e-3$\pm$5.15e-3     &  6105.0$\pm$907.9      &   102.5$\pm$25.2     &   13.6$\pm$1.7  \\
& \footnotesize WTA-I-10  &  9.72e-3$\pm$5.31e-3      &   5968.1$\pm$822.1     &    104.6$\pm$23.7    &  13.2$\pm$1.7 \\
& \footnotesize WTA-I-50 & 2.56e-2$\pm$3.82e-3      &  6337.5$\pm$693.2      &  78.1$\pm$22.9      &    19.0$\pm$1.3 \\
& \footnotesize WTA-I-100 &   3.14e-2$\pm$5.82e-3     &   8020.3$\pm$455.9     &    101.2$\pm$25.3    &   56.9$\pm$4.6 \\
& \footnotesize WTA-I-200 & 3.73e-2$\pm$1.10e-2     &    7530.2$\pm$321.6    &     131.8$\pm$11.1   &   82.1$\pm$8.8  \\

\midrule
\multirow{4}{*}{CIFAR10}
& \footnotesize WTA-I-1  &   2.15e-2$\pm$8.22e-3    &  8599.5$\pm$830.7     &  161.5$\pm$10.8    &  19.5$\pm$1.1    \\
& \footnotesize WTA-I-10 & 2.20e-2$\pm$8.13e-3     & 8401.2$\pm$928.3 &  162.2$\pm$10.3   &   19.3$\pm$1.1 \\
& \footnotesize WTA-I-50  &  4.09e-2$\pm$7.14e-3     & 6884.8$\pm$642.5  & 129.3$\pm$19.7    &  30.12$\pm$1.4 \\
& \footnotesize WTA-I-100 &  4.51e-2$\pm$8.14e-3    & 8500.2$\pm$625.1  & 136.2$\pm$22.1    &   68.9$\pm$4.5 \\
& \footnotesize WTA-I-200 &   5.38e-2$\pm$1.24e-2    & 8269.7$\pm$408.53  &  130.0$\pm$5.37   &  84.5$\pm$10.1  \\

\midrule
\multirow{4}{*}{ORL} & \footnotesize WTA-I-1  &  2.91e-2$\pm$6.07e-3    &  7320.0$\pm$847.0    &  169.4$\pm$11.7   &  19.5$\pm$1.8  \\ 
& \footnotesize WTA-I-10   &   3.06e-2$\pm$6.19e-3   &  7461.8$\pm$745.6    &  169.5$\pm$12.1  &  19.2$\pm$1.8  \\ 
& \footnotesize WTA-I-50   &   3.48e-2$\pm$6.48e-3   &  6254.8$\pm$716.5    & 124.7$\pm$27.3   &  31.7$\pm$1.8  \\
& \footnotesize WTA-I-100   &  3.90e-2$\pm$6.10e-3   &  8643.1$\pm$517.1    & 134.3$\pm$29.9   &  72.2$\pm$4.2  \\
& \footnotesize WTA-I-200  &   4.16e-2$\pm$4.90e-3   &  8107.6$\pm$405.9    & 134.9$\pm$3.13   &  90.1$\pm$11.6  \\
\bottomrule
\end{tabular}
\end{adjustbox}
\end{table}


\begin{figure}[!t]
\centering
\begin{tabular}{c c c c c c c c }
& {\footnotesize LGN} & & & & & \\
& {\footnotesize Target}  & {\footnotesize \textbf{ WTA-I-1} } & {\footnotesize WTA-I-10} & {\footnotesize WTA-I-50} & {\footnotesize WTA-I-100} & {\footnotesize WTA-I-200}\\
\rotatebox[origin=c]{90}{ {\footnotesize
 ORL }} & \raisebox{-0.5\height}{\includegraphics[width=0.10\textwidth]{figures/Figure5d.png}} & 
\raisebox{-0.5\height}{\includegraphics[width=0.10\textwidth, cfbox=black 2pt 1pt]{figures/Figure5h.png}} &
\raisebox{-0.5\height}{\includegraphics[width=0.10\textwidth]{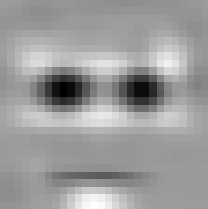}} & \raisebox{-0.5\height}{\includegraphics[width=0.10\textwidth]{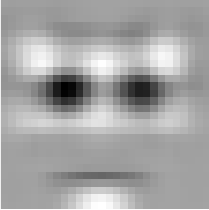}} & \raisebox{-0.5\height}{\includegraphics[width=0.10\textwidth]{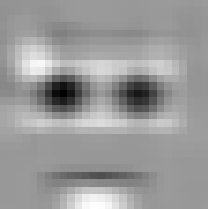}}
& \raisebox{-0.5\height}{\includegraphics[width=0.10\textwidth]{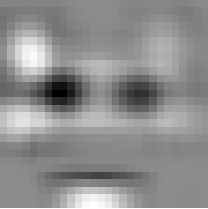}} \\
&  & {\footnotesize 0.03772} & {\footnotesize 0.04585 } & {\footnotesize 0.04995} & {\footnotesize 0.05747} & {\footnotesize 0.06470} \\

\rotatebox[origin=c]{90}{ {\footnotesize
 MNIST }} & \raisebox{-0.5\height}{\includegraphics[width=0.10\textwidth]{figures/Figure5i.png}} & 
\raisebox{-0.5\height}{\includegraphics[width=0.10\textwidth, cfbox=black 2pt 1pt]{figures/Figure5m.png}} & 
\raisebox{-0.5\height}{\includegraphics[width=0.10\textwidth]{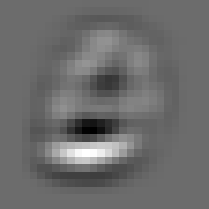}} & 
\raisebox{-0.5\height}{\includegraphics[width=0.10\textwidth]{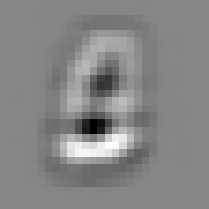}} & 
\raisebox{-0.5\height}{\includegraphics[width=0.10\textwidth]{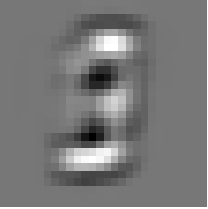}} & 
\raisebox{-0.5\height}{\includegraphics[width=0.10\textwidth]{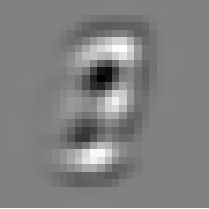}} \\
&  & {\footnotesize 0.01057} & {\footnotesize 0.02179} & {\footnotesize 0.03619} & {\footnotesize 0.03665} & {\footnotesize 0.04171} \\

\rotatebox[origin=c]{90}{ {\footnotesize
 F-MNIST }} & \raisebox{-0.5\height}{\includegraphics[width=0.10\textwidth]{figures/Figure5q.png}} & 
\raisebox{-0.5\height}{\includegraphics[width=0.10\textwidth, cfbox=black 2pt 1pt]{figures/Figure5u.png}} & 
\raisebox{-0.5\height}{\includegraphics[width=0.10\textwidth]{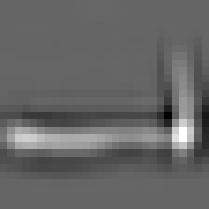}} & 
\raisebox{-0.5\height}{\includegraphics[width=0.10\textwidth]{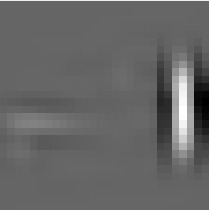}} & 
\raisebox{-0.5\height}{\includegraphics[width=0.10\textwidth]{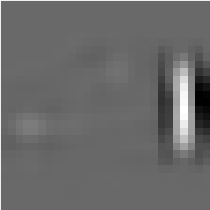}} & 
\raisebox{-0.5\height}{\includegraphics[width=0.10\textwidth]{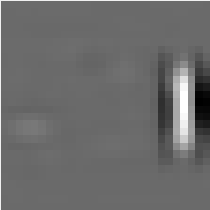}} \\
&  & {\footnotesize 0.01019} & {\footnotesize 0.01625} & {\footnotesize 0.02291} & {\footnotesize 0.02705} & {\footnotesize 0.03011} \\

\rotatebox[origin=c]{90}{ {\footnotesize
 CIFAR10 }} & \raisebox{-0.5\height}{\includegraphics[width=0.10\textwidth]{figures/Figure5aa.png}} & 
\raisebox{-0.5\height}{\includegraphics[width=0.10\textwidth, cfbox=black 2pt 1pt]{figures/Figure5ee.png}} & 
\raisebox{-0.5\height}{\includegraphics[width=0.10\textwidth]{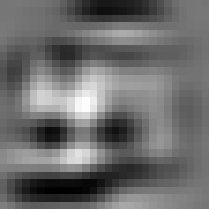}} & 
\raisebox{-0.5\height}{\includegraphics[width=0.10\textwidth]{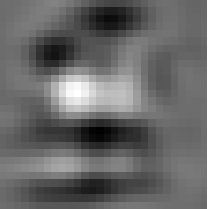}} & 
\raisebox{-0.5\height}{\includegraphics[width=0.10\textwidth]{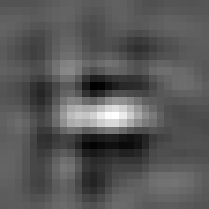}} & 
\raisebox{-0.5\height}{\includegraphics[width=0.10\textwidth]{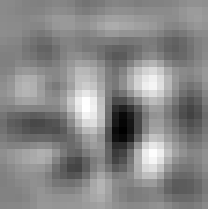}} \\
&  & {\footnotesize 0.02068} & {\footnotesize 0.03972} & {\footnotesize 0.04579} & {\footnotesize 0.05277} & {\footnotesize 0.05687} \\

\end{tabular}
\caption{Object representation using different WTA-I schemes, where between 1 (harder WTA-I 1) and 200 (softer WTA-I 200) neurons were active for each training sample. The number below each image indicates the reconstruction error for that particular image. Target and prediction images were normalized in $[0,1]$. The black frame highlights the image with the smallest error in each row.}
\label{fig:reconstructedImagesWTAI}
\end{figure}

\subsection{Winner-take-all inhibition}

\noindent We used a hard WTA-I scheme such that, if any V1 neuron fired during a certain iteration, it simultaneously prevented other neurons from firing until the next sample \citep{maass2000computational}. 
This scheme computes a function WTA-I$_n$: 
$\mathbb{R}^{n}\rightarrow \lbrace 0, 1 \rbrace^n$  
whose output 
$\langle y_1, \ldots, y_n \rangle = $ 
WTA-I$_n$ ($x_1,..., x_n$) satisfied:

\begin{equation}
y_i =\begin{cases} 1, \,\,\,\, $if  $x_i$ $>$ $x_j$ for all $j \neq i$ $\\
                     0, \,\,\,\, $otherwise$. 
       \end{cases} \label{eq:3}
\end{equation}

For a given set of $n$ different inputs $x_{1}, \ldots, x_{n}$, a hard WTA-I scheme would thus yield a single output $y_{i}$ with value $1$ (corresponding to the neuron that received the largest input $x_{i}$), whereas all other neurons would be silent. 
\citet{sanchez2022efficient} showed that a hard WTA-I scheme was essential for enforcing competition among neurons, which led to sparser object representations and lower reconstruction error compared to softer WTA-I schemes.

\subsection{Stimulus reconstruction}

\noindent The activity map $\xi_{j}$ of the $i$-th V1 neuron was estimated as follows:
\begin{equation}
\xi_{j} \approx  \sum\limits_{j\in{LGN}}w_{ij}\psi_{j},
\label{eq:4}
\end{equation}
where $\psi_{j}$ was the RF of the $j$-th LGN afferent, and $w_{ij}$ was the weight of the synapse connecting the $j$-th afferent to the $i$-th V1 neuron. 

Stimuli $k$ were then linearly reconstructed from the V1 population activity:
\begin{equation}
OR_{k}=  \sum\limits_{j\in{V1}}r_{kj}\xi_{j},
\label{eq:5}
\end{equation}
where $r_{kj}$ was the response of the $j$-th V1 neuron to the $k$-th image and $\xi_{j}$ was its activity map. Reconstruction error for an image $k$ was calculated as the pixel-wise mean square error between the LGN ($LGN_k$) and the V1 activity maps $OR_k$.

\subsection{Sparsity}

We computed a sparsity metric for the population activity in the network schemes according to the definition of sparsity by \citet{vinje2000sparse}. On average, we measured how many neurons were activated by any given stimulus (population sparsity) and for all active neurons, how many stimuli any given neuron responded to (lifetime sparsity), as can be seen in Equation~\ref{eq:s}). 

\begin{equation}
\mathrm{sparsity} =  \left( 1 - \frac{1}{N} \frac{(\sum_{n=1}r_{i})^{2}}{\sum_{n=1}r_{i}^{2}}  \right)  \bigg/ \left( 1 - \frac{1}{N} \right) ,
\label{eq:s}
\end{equation}

For population sparsity, $r_i$ was the response of the $i$-th neuron to a particular stimulus, and $N$ was the number of model neurons. For lifetime sparsity, $r_i$ was the response of a neuron to the $i$-th stimulus, and $N$ was the number of stimuli. Population sparsity was averaged across stimuli, and lifetime sparsity was averaged across neurons \citep{beyeler20163d}. We also calculated the average number of spikes per stimulus.

\subsection{Dataset}

\noindent To demonstrate the generality of our approach, we assessed the ability of our SNN network to represent visual stimuli from the MNIST \citep{lecun1998mnist}, FASHION-MNIST \citep{xiao2017fashion}, CIFAR10 \citep{krizhevsky2009learning} and ORL \citep{samaria1994parameterisation} datasets. MNIST is a dataset of handwritten digits and consists of 60,000 training patterns and 10,000 test patterns. FASHION-MNIST is a dataset of Zalando article images consisting of a training set of 60,000 examples and a test set of 10,000 examples. Each example of both, MNIST and FASHION-MNIST, is a $28 \times 28$ grayscale image, associated with a label from 10 classes. The CIFAR10 database consists of 60,000 $32 \times 32$ colour images in 10 classes, with 6,000 images per class. There are 50,000 training images and 10,000 test images. The ORL database of faces contains 400 images from 40 distinct subjects. The size of each image is $92 \times 112$ pixels, with 256 grey levels per pixel. 

We enlarged images from the CIFAR10 and ORL database using data augmentation with different orientations of the original images to match the data size with MNIST and FASHION-MNIST datasets.

\subsection{Statistical analysis}

Data were analyzed using two-way ANOVA and post hoc-test with Tukey’s method to evaluate simultaneously the effect of the two grouping variables (Dataset and Networks/WTA-I schemes/V1 neurons) on the following response variables: reconstruction error, spike count/neuron, lifetime sparsity, population sparsity, and recognition time with $*** = p<.001$; $** = p<.01$; $* = p<.05$ and $ns = p\ge.05$.

\section{Results}\label{sec4}

\subsection{Object representation using multi-scale network} 

\noindent The performance using a single-scale (i.e., low-scale, medium-scale, or high-scale networks) and multi-scale network is summarized in Fig.~\ref{fig:graph1}. 
The results show the reconstruction error, lifetime sparsity, population sparsity and spike count per neuron (mean $\pm$ standard deviation) achieved on the test sets for all databases.
The reconstruction error for the four networks (low-scale, medium-scale, high-scale and multi-scale) is shown in Fig.~\ref{fig:graph1}a.
We found similarity between the reconstruction errors of the three single networks (low, medium and high-scale) for all datasets, with some slight discrepancy in the more complex CIFAR10 and ORL datasets. Interestingly, the use of multi-scale manages to further reduce the reconstruction error, being the same trend for all datasets.
We also performed a test to determine if the mean difference between networks are statically significant using two-tailed test with a significant level $\alpha=0.05$.
The analysis of the average reconstruction error reveals a significant difference between networks (Low/Multi-scale, Medium/Multi-scale and High/Multi-scale). Examples of object representations for all datasets can be found in Fig.~\ref{fig:reconstructedImages}.

Fig.~\ref{fig:graph1}b shows the number of spikes per neuron needed for object representation. The number of spikes needed to represent an object decreased with the Multi-scale scheme compared to low, medium and high-scale networks. On the other hand, we found that the CIFAR10 and ORL dataset, which we considered two of the most complex of the four datasets, needed the highest number of spikes per neuron for all networks. 

Fig.~\ref{fig:graph1}c shows the number of distinct stimuli the neuron responds to during the lifetime of a neuron. The Multi-scale network showed a higher number of active stimuli for all datasets compared to the single networks. Moreover, we found significant differences between the networks, being more significant for Medium/Multi-scale and High/Multi-scale. The same trend was found for the population sparsity, where the Multi-scale presented more active neurons than the low, medium and high-scale networks and significant differences were found between them (see Fig.~\ref{fig:graph1}d).




\begin{figure}[!t]
     \centering
     \begin{subfigure}[b]{0.45\textwidth}
         \centering
         \includegraphics[width=\textwidth]{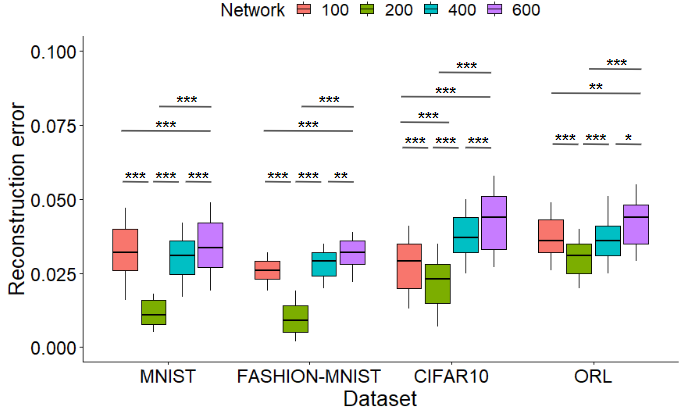}
         \caption[b]{}
     \end{subfigure}
    \begin{subfigure}[b]{0.45\textwidth}
         \centering
         \includegraphics[width=\textwidth]{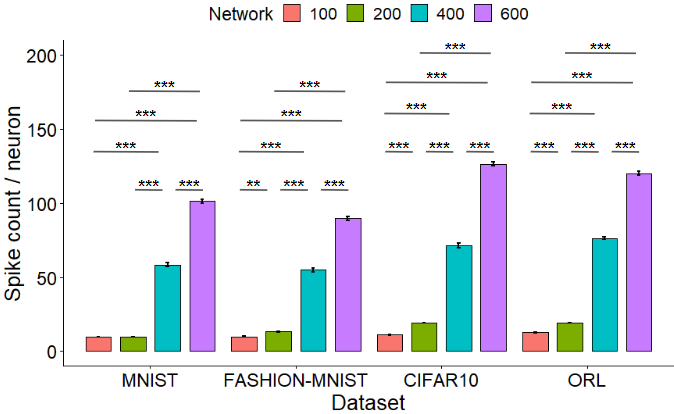}
         \caption[b]{}
     \end{subfigure}
     \begin{subfigure}[b]{0.45\textwidth}
         \centering
         \includegraphics[width=\textwidth]{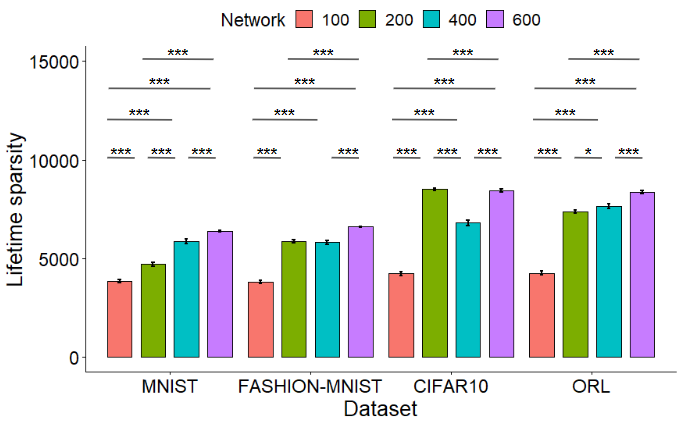}
         \caption[b]{}
     \end{subfigure}
    \begin{subfigure}[b]{0.45\textwidth}
         \centering
         \includegraphics[width=\textwidth]{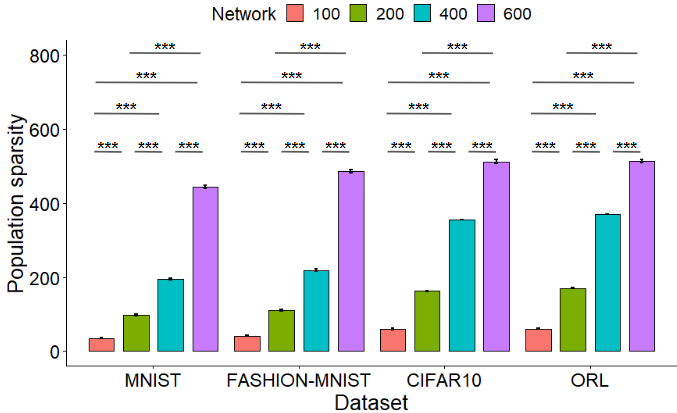}
         \caption[b]{}
     \end{subfigure}
        \caption{V1 neurons. (a) Reconstruction error of test set using different number of V1 neurons: 100, 200, 400 and 600. (b) Lifetime sparsity: active stimuli during the lifetime of a neuron. (c) Population sparsity: neurons active at any point in time. (d) Spike count per neuron: number of spikes fired by an active neuron. Mean responses and standard deviation grouped by type of network architecture (Low-scale, Medium-scale, High-scale and Multi-scale). Error bars have been averaged across neurons for lifetime sparsity and averaged across images for population sparsity.}
        \label{fig:graph2}
\end{figure}

\begin{figure}[!t]
\centering
\begin{tabular}{c c c c c c c }
& {\footnotesize LGN} & & & & \\
& {\footnotesize Target} & {\footnotesize 100} & {\footnotesize 200} & {\footnotesize 400 } & {\footnotesize 600} \\

\rotatebox[origin=c]{90}{{\footnotesize ORL}} & \raisebox{-0.5\height}{\includegraphics[width=0.12\textwidth]{figures/Figure5a.png}} &
\raisebox{-0.5\height}{\includegraphics[width=0.12\textwidth]{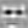}} &
\raisebox{-0.5\height}{\includegraphics[width=0.12\textwidth, cfbox=black 2pt 1pt]{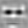}} &
\raisebox{-0.5\height}{\includegraphics[width=0.12\textwidth]{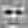}} &
\raisebox{-0.5\height}{\includegraphics[width=0.12\textwidth]{figures/V1neurons/orl/400.png}}\\
& & {\footnotesize 0.04945} & {\footnotesize 0.03772} & {\footnotesize 0.04956} & {\footnotesize 0.05407} \\

\rotatebox[origin=c]{90}{ {\footnotesize
 MNIST}} & \raisebox{-0.5\height}{\includegraphics[width=0.12\textwidth]{figures/Figure5i.png}} &
\raisebox{-0.5\height}{\includegraphics[width=0.12\textwidth]{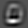}} & \raisebox{-0.5\height}{\includegraphics[width=0.12\textwidth,cfbox=black 2pt 1pt]{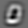}} & \raisebox{-0.5\height}{\includegraphics[width=0.12\textwidth]{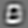}} & \raisebox{-0.5\height}{\includegraphics[width=0.12\textwidth]{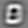}} \\
& & {\footnotesize 0.02702} & {\footnotesize 0.01057} & {\footnotesize 0.04166} & {\footnotesize 0.04719} \\

\rotatebox[origin=c]{90}{{\footnotesize F-MNIST}} & \raisebox{-0.5\height}{\includegraphics[width=0.12\textwidth]{figures/Figure5q.png}} &
\raisebox{-0.5\height}{\includegraphics[width=0.12\textwidth]{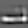}} &
\raisebox{-0.5\height}{\includegraphics[width=0.12\textwidth,cfbox=black 2pt 1pt]{figures/Figure5x.png}} &
\raisebox{-0.5\height}{\includegraphics[width=0.12\textwidth]{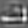}} &
\raisebox{-0.5\height}{\includegraphics[width=0.12\textwidth]{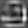}}\\
& & {\footnotesize 0.02597} & {\footnotesize 0.01019} & {\footnotesize 0.02740} & {\footnotesize 0.03268} \\

\rotatebox[origin=c]{90}{{\footnotesize CIFAR10}} & \raisebox{-0.5\height}{\includegraphics[width=0.12\textwidth]{figures/Figure5aa.png}} &
\raisebox{-0.5\height}{\includegraphics[width=0.12\textwidth]{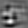}} &
\raisebox{-0.5\height}{\includegraphics[width=0.12\textwidth, cfbox=black 2pt 1pt]{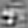}} &
\raisebox{-0.5\height}{\includegraphics[width=0.12\textwidth]{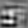}} &
\raisebox{-0.5\height}{\includegraphics[width=0.12\textwidth]{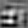}}\\
& & {\footnotesize 0.03164} & {\footnotesize 0.02067} & {\footnotesize 0.03639} & {\footnotesize 0.03924} \\

\end{tabular}
\caption{Object representation with Multi-scale network varying the number of V1 neurons: 100, 200, 400 and 600 neurons. The number below each image indicates the reconstruction error for that particular image. The black frame highlights the image with the smallest error.}
\label{fig:V1reconstructedImages}
\end{figure}



\begin{table}[]
\caption{Global results for V1 neurons. Comparison of mean responses and standard deviation grouped by type of V1 neurons and dataset.}\label{tab3}%
 \begin{adjustbox}{width=\textwidth}
\begin{tabular}{|c|c|c|c|c|c|}
\toprule
{Dataset} & {V1 neurons} & \footnotesize RE   & \footnotesize LS   & \footnotesize PS   & \footnotesize SC  \\ 
\midrule

\multirow{4}{*}{MNIST} & \footnotesize 100  &  2.62e-2$\pm$8.61e-3    &   3667.8$\pm$718.3   & 35.2$\pm$9.3     &  10.3$\pm$0.8   \\
& \footnotesize 200  &  1.16e-2$\pm$3.77e-3    &  4500.1$\pm$782.7    &  95$\pm$19.9    &   10$\pm$0.79  \\
& \footnotesize 400 & 2.94e-2$\pm$7.72e-3    &  5719.3$\pm$1568.7    &  206.2$\pm$9.3    &  59.7$\pm$13.1 \\
& \footnotesize 600 &   3.47e-2$\pm$8.29e-3   &   6379.4$\pm$388.3   & 445.4$\pm$38.7      &  101.8$\pm$13.4   \\

\midrule
\multirow{4}{*}{FASHION-MNIST} & \footnotesize 100 &  2.56e-2$\pm$4.38e-3    &  3893.7$\pm$694.5    &  43.1$\pm$10.2     &  9.9$\pm$1.5   \\
& \footnotesize 200 & 9.34e-3$\pm$5.15e-3     &  6105.0$\pm$907.9      &   102.5$\pm$25.2     &   13.6$\pm$1.76 \\
& \footnotesize 400 &  2.78e-2$\pm$4.48e-3    &  6251.3$\pm$979.1    &  216.7$\pm$31.8     &  53.4$\pm$11.9 \\
& \footnotesize 600 &   3.09e-2$\pm$5.45e-3   &   6585.8$\pm$279.6   & 486.5$\pm$37.1      &  87.8$\pm$13.3   \\

\midrule
\multirow{4}{*}{CIFAR10} & \footnotesize 100 &  3.15e-2$\pm$9.37e-3   & 4374.1$\pm$1173.1 &  61.2$\pm$18.3  &  11.1$\pm$1.4   \\
& \footnotesize 200 & 2.15e-2$\pm$8.22e-3    &  8599.5$\pm$830.7     &  161.5$\pm$10.8    &  19.5$\pm$1.06 \\
& \footnotesize 400 &  3.71e-2$\pm$7.60e-3   & 6498.8$\pm$1005.8 & 354.0$\pm$29.2   &  70.3$\pm$12.3   \\
& \footnotesize 600 &   4.37e-2$\pm$1.01e-2  & 8404.8$\pm$876.3 &  498.3$\pm$41.8  &  126.1$\pm$12.1  \\

\midrule
\multirow{4}{*}{ORL} & \footnotesize 100 &  3.73e-2$\pm$7.09e-3  & 4092.1$\pm$1058.3    &  58.7$\pm$14.9 &  12.4$\pm$1.7  \\ 
& \footnotesize 200    &    2.91e-2$\pm$6.07e-3    &  7320$\pm$847.0    &  169.4$\pm$11.7   &  19.5$\pm$1.78  \\ 
& \footnotesize 400   &   3.90e-2$\pm$7.90e-3   &  7769.2$\pm$1151.9   & 370.5$\pm$9.0  &  75.5$\pm$8.8  \\ 
& \footnotesize 600   &    4.17e-2$\pm$7.70e-3  &  8288.6$\pm$828.0   & 513.7$\pm$45.3  &  120.4$\pm$12.5  \\ 

\bottomrule
\end{tabular}
\end{adjustbox}
\end{table}

\subsection{Object representation using multi-scale network with varying number of V1 neurons} 

\noindent Fig.~\ref{fig:graph2} (a) shows the reconstruction error after training for the test set using different numbers of V1 neurons. We found that the reconstruction error went through a minimum (at roughly 200 V1 neurons) for all databases, 
which is consistent with the bias-variance dilemma \citep{beyeler2019neural}. 
It seems that using a larger number of neurons with our multi-scale network leads to overfitting and a less sharp reconstruction, as can be seen in Fig.~\ref{fig:V1reconstructedImages}.

In addition, the number of neurons needed to represent an object increased with the number of V1 neurons, nearly tripling the spikes from 200 to 400 neurons and quintupling from 200 to 600 (Fig.~\ref{fig:graph2}c). Increasing the V1 population beyond 200 neurons did therefore not lead to any visible benefits in reconstruction error (Fig.~\ref{fig:V1reconstructedImages}). We therefore limited our V1 population to 200 neurons for all subsequent simulations and analyses.

\subsection{Object representation using soft WTA-I schemes} 

\noindent We also tested object representation using various soft WTA-I schemes, where we varied the number of V1 neurons allowed to be active for each training image (see Fig.~\ref{fig:WTAISpikes}). Fig.~\ref{fig:WTAISpikes}a shows the reconstruction error on the test set across the range of possible WTA-I schemes, ranging from hard (where only one neuron was active per image) to soft (where all 200 neurons were active).

We found that the softer the WTA-I scheme, the higher the reconstruction error. The reason for this became evident when we visualized the resulting object representations (Fig.~\ref{fig:reconstructedImagesWTAI}). 
WTA-I schemes where at most 10 neurons were allowed to be active were instrumental in maintaining competition among neurons. In the absence of a strong WTA-I scheme, multiple neurons ended up learning similar visual features, which resulted in poor object reconstruction (right half of Fig.~\ref{fig:reconstructedImagesWTAI}). Also, due to this overlap between neurons, the final feature set was quite limited.

We also found that both the active stimuli during the lifetime of a neuron and the active neurons increased with the number of V1 neurons allowed to be active during training (see Fig.~\ref{fig:WTAISpikes}c, d). Furthermore, the number of spikes needed to represent an object showed the same trend (Fig.~\ref{fig:WTAISpikes}b).

\section{Discussion}\label{sec5}

\noindent In this work, we have proposed an SNN model that uses spike-latency coding and WTA-I to efficiently represent visual stimuli using multi-scale parallel processing.
In particular, this paper developed an extension of earlier work
\citep{chauhan2018emergence,chauhan2021sub,sanchez2022efficient}
to investigate how the quality of the represented objects changes under different schemes of the primary visual system processing with subsets of neurons tuned to different SF scales. 

We found that the multi-scale network outperformed all three single-scale networks across all datasets (Fig.~\ref{fig:graph1}), sacrificing sparsity for a lower reconstruction error.
However, it is interesting to note that the multi-scale network used the smallest average number of spikes per neuron (Fig.~\ref{fig:graph1}b) across all datasets, indicating that it favored a code where many neurons were weakly activated.
In all cases, the learned receptive fields (Fig.~\ref{fig:rfs}) were in agreement with nonnegative sparse coding (NSC), which is an efficient population coding scheme based on dimensionality reduction and sparsity constraints that promotes sparse and parts-based population codes \citep{beyeler2019neural}.


We also studied how the number of V1 neurons in the network affected reconstruction error and sparsity of the learned population code.
In agreement with previous work on NSC \citep{beyeler20163d,beyeler2019neural}, we found that the reconstruction error (on the test set) goes through a minimum as a function of network size (Fig.~\ref{fig:graph2}a).
This minimum is though to indicate the optimal model complexity according to the bias-variance dilemma; that is, the point at which the model's generalization error is minimized.
Curiously, this ``sweet spot'' was found to be at roughly 200 V1 neurons for all tested datasets
(Fig.~\ref{fig:V1reconstructedImages}).
On the other hand, sparsity increased monotonically with network size (Fig.~\ref{fig:graph2}b--d), which is more in line with the traditional sparse coding literature \citep{olshausen_sparse_1997}.

We also implemented various soft WTA-I schemes to investigate how the quality of represented objects changed (Fig.~\ref{fig:WTAISpikes}). The WTA-I soft schemes consisted of 10, 50, 100, and 200 (i.e., all) neurons firing during a given iteration, while all other neurons were silent.
We found that the softer the WTA-I scheme, the larger the reconstruction error (Fig.~\ref{fig:WTAISpikes}a) and the number of spikes needed to represent an object (Fig.~\ref{fig:WTAISpikes}b). The reason for this became clear when we visualized the resulting object representations (Fig.~\ref{fig:reconstructedImagesWTAI}). In the absence of a strong WTA-I scheme, multiple neurons ended up learning similar visual features, thus resulting in poor object reconstructions (Fig.~\ref{fig:reconstructedImagesWTAI}).

Although our network was able to efficiently represent images from various datasets, 
an important issue that we did not address in this paper is a comparison with other SNNs with other forms of STDP (e.g., with an additive instead of a multiplicative rule) and/or to SNNs trained with other learning scheme (e.g., SNNs trained with the surrogate gradient).
In addition, a future extension of the model might focus on deeper architectures with parallel processing with multiple scales and more challenging visual stimuli.


\section{Conclusion}

In conclusion, we have shown that a network of spiking neurons 
tuned to different SFs
can represent objects with as little as 15 spikes per neuron using spike-latency coding and WTA-I.
WTA-I schemes were essential for enforcing competition among neurons, which led to sparser object representations and lower reconstruction errors.
Studying how object recognition may be implemented using biologically plausible learning rules in SNNs may not only further our understanding of the brain, but also lead to new efficient artificial vision systems.






\section*{Acknowledgments}

This work was partially supported by a UCSB Academic Senate Faculty Research Grant to MB and by FLAG-ERA project JTC-2019 DOMINO to BRC. TC was supported by a fellowship from the JPB Foundation, and grant FRM:SPF20170938752 from the Fondation pour la Recherche Médicale.


\section*{Author Contributions}

TC and BRC conceived and designed the original study, which was subsequently extended by MSG and MB. 
TC wrote all the code and MSG ran all the simulations. 
MSG and MB analyzed and interpreted the results. 
MSG drafted the manuscript. 
All authors reviewed and approved the final version of the manuscript.

\clearpage
\begin{appendices}

\section{Comparison between Multi-scale and Lateral-scale network architectures}\label{secA1}

\begin{figure}[b]
    \centering
    \includegraphics[width=0.8\linewidth]{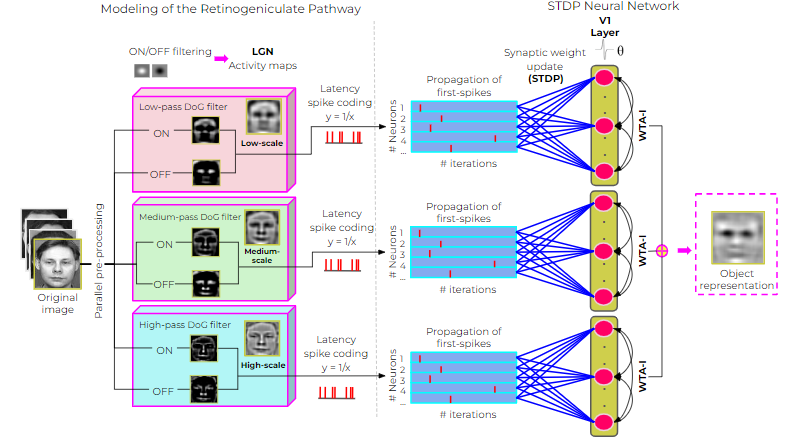}
    \caption{Lateral-scale network. Images from the ORL dataset \citep{samaria1994parameterisation} were convolved with ON and OFF center-surround kernels to simulate responses in the LGN. We used three LGN sub-networks processed based on a particular SF: Low-scale, Medium-scale and High-scale (see Fig.~\ref{fig:preprocessing}). The three LGN responses were converted to spike latencies and fed to a SNN each, resulting in three lateral SNN with plastic synapses implementing STDP and WTA-I. The reconstructed images resulted of the three lateral networks were added at the end for the object reconstruction.}
    \label{fig:overview2}
\end{figure}

We propose another network architecture called `Lateral-scale' that also uses parallel processing of multiple scales (see Fig.~\ref{fig:overview2}). In this case, the LGN preprocessing is the same as in the Multi-scale network architecture, but now the three LGN responses were converted to spike latencies and fed to a SNN each, resulting in three lateral SNN with plastic synapses implementing STDP and WTA-I. The reconstructed images resulted of the three lateral sub-networks were added at the end of the training for the object representation.

\begin{figure}[!t]
     \centering
     \begin{subfigure}[b]{0.45\textwidth}
         \centering
         \includegraphics[width=\textwidth]{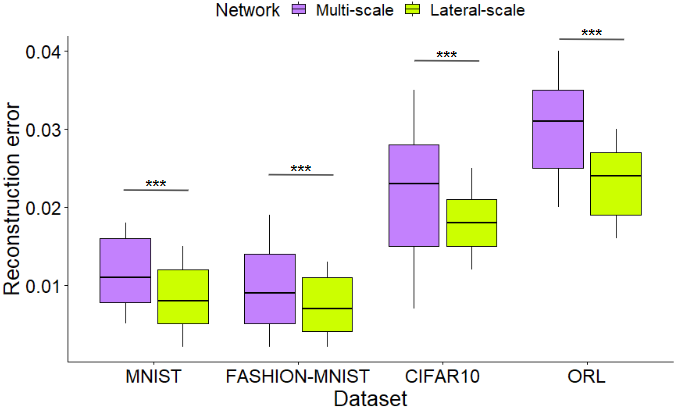}
         \caption[a]{}
     \end{subfigure}
     \begin{subfigure}[b]{0.45\textwidth}
         \centering
         \includegraphics[width=\textwidth]{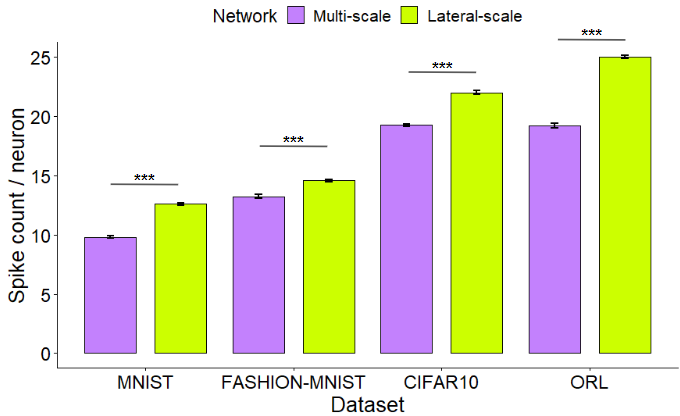}
         \caption{}
     \end{subfigure}
     \begin{subfigure}[b]{0.45\textwidth}
         \centering
         \includegraphics[width=\textwidth]{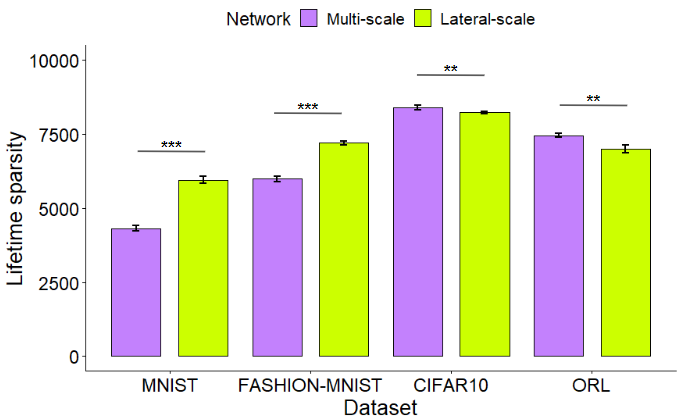}
         \caption{}
     \end{subfigure}
     \begin{subfigure}[b]{0.45\textwidth}
         \centering
         \includegraphics[width=\textwidth]{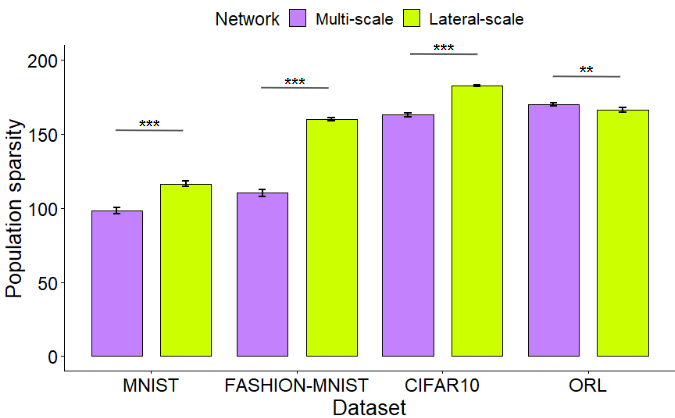}
         \caption{}
     \end{subfigure}
        \caption{(a) Reconstruction error of test set using Multi-scale and Lateral-scale networks. (b) Number of spikes per neuron needed for the object representation using Multi-scale and Lateral-scale networks. (c) Lifetime sparsity: active stimuli during the lifetime of a neuron. (d) Population sparsity: neurons active at any point in time. $*** = p<.001$; $** = p<.01$; $* = p<.05$; $ns = p>.05$. All t-tests paired samples, two-tailed.} 
        \label{fig:graphA2}
\end{figure}

\begin{figure}[!t]
\centering
\begin{tabular}{c c c c c c c c}
& {\footnotesize LGN} & {\footnotesize Multi-scale} & {\footnotesize Lateral-scale} & {\footnotesize LGN} &{\footnotesize Multi-scale} & {\footnotesize Lateral-scale} \\
& {\footnotesize Target A} & {\footnotesize A} & {\footnotesize A} & {\footnotesize Target B} &{\footnotesize B} & {\footnotesize B} \\

\rotatebox[origin=c]{90}{ {\footnotesize
 ORL}} & \raisebox{-0.5\height}{\includegraphics[width=0.10\textwidth]{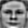}} &
\raisebox{-0.5\height}{\includegraphics[width=0.10\textwidth]{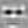}} & \raisebox{-0.5\height}{\includegraphics[width=0.10\textwidth,cfbox=black 2pt 1pt]{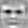}} & \raisebox{-0.5\height}{\includegraphics[width=0.10\textwidth]{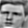}} & \raisebox{-0.5\height}{\includegraphics[width=0.10\textwidth]{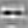}} & \raisebox{-0.5\height}{\includegraphics[width=0.10\textwidth,cfbox=black 2pt 1pt]{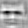}} \\
& & {\footnotesize 0.03772} & {\footnotesize 0.02810} & &{\footnotesize 0.02769} & {\footnotesize 0.01575} \\

\rotatebox[origin=c]{90}{ {\footnotesize
 MNIST}} & \raisebox{-0.5\height}{\includegraphics[width=0.10\textwidth]{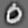}} &
\raisebox{-0.5\height}{\includegraphics[width=0.10\textwidth]{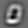}} & \raisebox{-0.5\height}{\includegraphics[width=0.10\textwidth,cfbox=black 2pt 1pt]{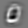}} & \raisebox{-0.5\height}{\includegraphics[width=0.10\textwidth]{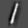}} & \raisebox{-0.5\height}{\includegraphics[width=0.10\textwidth]{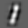}} & \raisebox{-0.5\height}{\includegraphics[width=0.10\textwidth,cfbox=black 2pt 1pt]{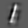}} \\
& & {\footnotesize 0.01057} & {\footnotesize 0.00942} & &{\footnotesize 0.01813} & {\footnotesize 0.01275} \\

\rotatebox[origin=c]{90}{ {\footnotesize
 F-MNIST}} & \raisebox{-0.5\height}{\includegraphics[width=0.10\textwidth]{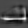}} &
\raisebox{-0.5\height}{\includegraphics[width=0.10\textwidth]{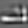}} & \raisebox{-0.5\height}{\includegraphics[width=0.10\textwidth,cfbox=black 2pt 1pt]{figures/Apendix/fashion_mnist/lateral1.png}} & \raisebox{-0.5\height}{\includegraphics[width=0.10\textwidth]{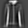}} & \raisebox{-0.5\height}{\includegraphics[width=0.10\textwidth]{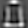}} & \raisebox{-0.5\height}{\includegraphics[width=0.10\textwidth,cfbox=black 2pt 1pt]{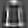}} \\
& & {\footnotesize 0.01019} & {\footnotesize 0.00936} & &{\footnotesize 0.01419} & {\footnotesize 0.01029} \\
\rotatebox[origin=c]{90}{ {\footnotesize
 CIFAR10}} & \raisebox{-0.5\height}{\includegraphics[width=0.10\textwidth]{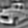}} &
\raisebox{-0.5\height}{\includegraphics[width=0.10\textwidth]{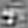}} & \raisebox{-0.5\height}{\includegraphics[width=0.10\textwidth,cfbox=black 2pt 1pt]{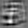}} & \raisebox{-0.5\height}{\includegraphics[width=0.10\textwidth]{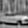}} & \raisebox{-0.5\height}{\includegraphics[width=0.10\textwidth]{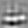}} & \raisebox{-0.5\height}{\includegraphics[width=0.10\textwidth,cfbox=black 2pt 1pt]{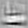}} \\
& & {\footnotesize 0.02067} & {\footnotesize 0.01673} & &{\footnotesize 0.02884} & {\footnotesize 0.02423} \\

\end{tabular}
\caption{Object representation for Multi-scale and Lateral-scale network architectures using 200 V1 neurons. Two examples of object representation (image A and image B) for Multi-scale and Lateral-scale architectures and for the four databases. The Lateral-scale scheme recognizes some finer details in the image compared to Multi-scale, where the image details are coarser. The number below each image indicates the reconstruction error for that particular image. The black frame highlights the image with the smallest error.}
\label{fig:Multi-scalevsLateral-scale}
\end{figure}

As shown in Fig.~\ref{fig:graphA2}a, the Lateral-scale network results in a lower but very similar reconstruction error than the proposed Multi-scale network. This may be because the Lateral-scale scheme recognizes a few more details corresponding to fine details in the image (see Fig.~\ref{fig:Multi-scalevsLateral-scale}). Lateral-scale was not significantly better than Multi-scale if we refer to the representation of objects (see Fig.~\ref{fig:Multi-scalevsLateral-scale} but used significantly more spikes (Fig.~\ref{fig:Multi-scalevsLateral-scale}b). The number of spikes required for reconstruction increases by approximately double spikes/neuron in some datasets. One drawback in Lateral-scale network is that we are training three lateral sub-networks, that means three times more trainable weights.

\begin{table}[]
\caption{Global results for Multi and Lateral-scale. Comparison of mean responses and standard deviation grouped by type of Multi and Lateral-scale and dataset.}\label{tab7}%
 \begin{adjustbox}{width=\textwidth}
\begin{tabular}{|c|c|c|c|c|c|}
\toprule
{Dataset} & {Network} & \footnotesize RE   & \footnotesize LS   & \footnotesize PS   & \footnotesize SC  \\ 
\midrule

\multirow{2}{*}{MNIST} & \footnotesize  Multi-scale   &  1.16e-2$\pm$3.77e-3    &  4500.1$\pm$782.7    &  95.0$\pm$19.9    &   10.0$\pm$0.79    \\
& \footnotesize  Lateral-scale  & 8.27e-3$\pm$4.04e-3    &   5867.9$\pm$444.1   &  115.6$\pm$4.9    &  12.5$\pm$1.4   \\

\midrule
\multirow{2}{*}{FASHION-MNIST} & \footnotesize  Multi-scale &  9.34e-3$\pm$5.15e-3     &  6105.0$\pm$907.9      &   102.5$\pm$25.2     &   13.6$\pm$1.76 \\
& \footnotesize  Lateral-scale & 6.93e-3$\pm$3.53e-3    &  7160.4$\pm$745.1   & 161.2$\pm$11.5   &  14.4$\pm$1.2   \\

\midrule
\multirow{2}{*}{CIFAR10} & \footnotesize  Multi-scale &  2.15e-2$\pm$8.22e-3    &  8599.5$\pm$830.7     &  161.5$\pm$10.8    &  19.5$\pm$1.06 \\
& \footnotesize  Lateral-scale &  1.87e-2$\pm$4.23e-3   &  8258.7$\pm$1130.5   &  182.4$\pm$19.2  & 21.9$\pm$1.2   \\

\midrule
\multirow{2}{*}{ORL} & \footnotesize Multi-scale   &  2.91e-2$\pm$6.07e-3    &  7320$\pm$847.0    &  169.4$\pm$11.7   &  19.5$\pm$1.78  \\
& \footnotesize Lateral-scale  & 2.28e-2$\pm$4.84e-3   &  7233.0$\pm$1241.5  &  168.3$\pm$17.4  & 25.1$\pm$1.5  \\
\bottomrule
\end{tabular}
\end{adjustbox}
\end{table}

\end{appendices}


\clearpage
\bibliography{sn-bibliography}


\end{document}